\documentclass{nature}

\usepackage{graphicx}
\usepackage{epsfig}
\usepackage{amssymb}

\def\apj{Astroph. J}

\def\aap{Astron. Astrophys.}

\def\mnras{Mon. Not. R. Astron. Soc.}

\title{Discrete sources as the origin of the Galactic X-ray ridge emission}
\author{Revnivtsev M.$^{1,2}$, Sazonov S.$^{2,3}$, Churazov E. $^{3,2}$, 
Forman W. $^{4}$, Vikhlinin A.$^{4,2}$, Sunyaev R.$^{3,2}$}
\begin{document}
\maketitle
\begin{affiliations}
 \item Excellence Cluster Universe, Technische Universit\"at M\"unchen, Garching, Germany
 \item Space Research Institute, Moscow, Russia
 \item Max-Planck-Instutut fuer Astrophysik, Garching, Germany
 \item Harvard-Smithsonian Center for Astrophysics, Cambridge, USA
\end{affiliations}

\begin{abstract}
An unresolved X-ray glow (at energies above a few kiloelectronvolts)
was discovered about 25 years ago and found to be coincident
with the Galactic disk—the Galactic ridge X-ray emission\cite{worrall82,warwick85}.
This emission \cite{koyama86,yamauchi93,yamauchi96,hands04,ebisawa05,koyama07,ebisawa08,yamauchi08} has a spectrum characteristic of a $10^8$K optically
thin thermal plasma, with a prominent iron emission line at
6.7 keV. The gravitational well of the Galactic disk, however, is far
too shallow to confine such a hot interstellar medium; instead, it
would flow away at a velocity of a few thousand kilometres per
second, exceeding the speed of sound in gas. To replenish the
energy losses requires a source of $10^{43}$ erg s$^{-1}$, exceeding by orders
of magnitude all plausible energy sources in the Milky Way\cite{tanaka02}. An
alternative is that the hot plasma is bound to a multitude of faint
sources\cite{worrall83}, which is supported by the recently observed similarities
in the X-ray and near-infrared surface brightness distributions\cite{mikej06,revnivtsev06}
(the latter traces the Galactic stellar distribution). Here we report
that at energies of 6-7 keV, more than 80 per cent of the seemingly
diffuse X-ray emission is resolved into discrete sources,
probably accreting white dwarfs and coronally active stars.
\end{abstract}

Observations clearly show that some fraction of the X-ray emission
of the Galaxy is produced by hot truly diffuse interstellar plasma,
heated by e.g. supernovae
\cite{park04}, while the bulk of previously unresolved X-ray emission
at energies above 1--2 keV remained unexplained. The strong
similarity of the Galactic ridge X-ray emission (GRXE) large scale distribution and that of the NIR
map of the Milky Way suggested a stellar origin of this emission. The
stellar origin was further supported by the close agreement between
the X-ray emissivity per unit stellar mass inferred for the GRXE and
the collective X-ray emissivity of the stellar population
within a few hundred parsecs of the Sun\cite{sazonov06}.

These findings motivated us to
perform in 2008 a decisive test with an ultra-deep, 1~Msec,
observation of a small ($\sim 16\times 16$ arcmin) field near the
Galactic Centre ($l^{II}=0.08$, $b^{II}=-1.42$) with the \emph{Chandra
  X-ray Observatory}.  We selected this region of the Galactic plane
because, here, a high GRXE intensity (essential for minimizing
the contribution from extragalactic sources)
combines with weak interstellar absorption (crucial for
maximizing the 0.5--7 keV \emph{Chandra} sensitivity for discrete sources). 
From what we know about the Solar neighbourhood, we can expect the
sources producing the 
bulk of the GRXE to be as faint as $\sim 10^{30}$~erg~s$^{-1}$ and to
have a surface density of $10^5$ per sq. deq or even higher in the
Galactic plane. Only with the combination of an ultra-deep
exposure and the excellent angular resolution of \emph{Chandra}
($\sim0.5$ arcsec \cite{weisskopf02}) has the task of resolving the
GRXE become possible.

To place the most stringent limits on the fraction of the GRXE
resolvable into discrete sources, we selected for our analysis a field
where i) the telescope's angular resolution is best and ii) spatial
variations of the soft X-ray emission below 1.5 keV (which might be
caused by supernova remnants) are minimal. We therefore restrict our
present study to a small circle of $2.56$ arcmin-radius near the telescope
optical axis (see Fig. 1). Below we refer to this field
($l^{II}=0.113$, $b^{II}=-1.424$) as "HRES" (stands for "High
Resolution").

The total measured X-ray surface brightness 
in HRES is $I_{\rm
  3-7~keV}=(4.6\pm0.4)\times10^{-11}$~erg ~s$^{-1}$ ~cm$^{-2}$ ~deg$^{-2}$ 
in the 3--7~keV band, or equivalently $I_{\rm 2-10~keV}=(8.6\pm0.5)\times 
10^{-11}$ erg ~s$^{-1}$ ~cm$^{-2}$ ~deg$^{-2}$ in the more conventional 
2--10~keV band, or, in scale-free units, 
$I_{\rm 2-10~keV}=3.8\pm0.2$ mCrab ~deg$^{-2}$ 
(here and below the uncertainties are 68\% confidence intervals
which include both statistical and count rate to flux conversion
uncertainties). 
The brightest source
detected in our region has a 2--10 keV flux $\sim1.8\times10^{-14}$
erg~s$^{-1}$~cm$^{-2}$ and thus a luminosity $\sim10^{32}$
erg~s$^{-1}$ if it is located at approximately the Galactic Centre
distance ($\sim 8$ kpc). More luminous, rarer sources are
found in our {\sl Chandra field}, but outside HRES; 
we exclude such sources ($L_{\rm
  2-10~keV}>10^{32}$ erg~s$^{-1}$) from consideration when addressing
the resolved fraction of the GRXE below.

The total measured X-ray surface
brightness must include the contribution from the nearly isotropic
extragalactic X-ray background (CXB\cite{giacconi62}). The mean CXB
intensity over the sky measured by \emph{Chandra} 
in the 2--10 keV energy band\cite{hickox06}
$I_{\rm CXB,~2-10~keV}= 2.19\times10^{-11}$
erg~s$^{-1}$~cm$^{-2}$~deg$^{-2}$, of which 31\% is provided\cite{moretti03} by
sources (mostly active galactic nuclei and quasars) brighter than
$2\times10^{-14}$ erg~s$^{-1}$~cm$^{-2}$. 
Given the absence of such bright sources in
HRES, the total CXB contribution is
$\sim1.5\times10^{-11}$ erg~s$^{-1}$~cm$^{-2}$~deg$^{-2}$. After 
subtraction of this extragalactic emission, the GRXE intensity in
HRES $I_{\rm
  GRXE,~2-10~keV}=(7.1\pm0.5)\times10^{-11}$~erg~s$^{-1}$~cm$^{-2}$~deg$^{-2}$. 

Looking at the same field in the near-infrared band, which provides
the best window on the Galactic stellar mass distribution, the 3.5$\mu m$
intensity measured with the {\sl Spitzer} IRAC instrument 
is $21\pm2$ MJy~sr$^{-1}$ (the uncertainty being mainly
due to the variance of the number of bright NIR point sources
within the small area of the study). Given the interstellar extinction
towards HRES 
$A_V\sim 3.5-4.5$ \cite{dutra03} and
adopting $A_{3.5~\mu m}/A_V=0.066$ \cite{dutra03,indebetouw05}, 
the extinction corrected NIR
surface brightness $I_{3.5~\mu m}=$ 26--29~MJy~sr$^{-1}$. Therefore,
the GRXE to NIR intensity ratio in HRES $I_{\rm
  2-10~keV}$ ($10^{-11}$ erg~s$^{-1}$~cm$^{-2}$~deg$^{-2}$)/$I_{\rm
  3.5~\mu m}$ (MJy~sr$^{-1}$) $=0.25\pm0.04$, in perfect agreement
with the value characterizing the entire Galaxy, $0.26\pm0.05$, deduced
from large-scale mapping of the GRXE\cite{mikej06}. This confirms that
the findings of the present study of a tiny region of the Galaxy may
be regarded as representative of the GRXE as a whole.

\begin{figure}
\includegraphics[width=0.5\columnwidth]{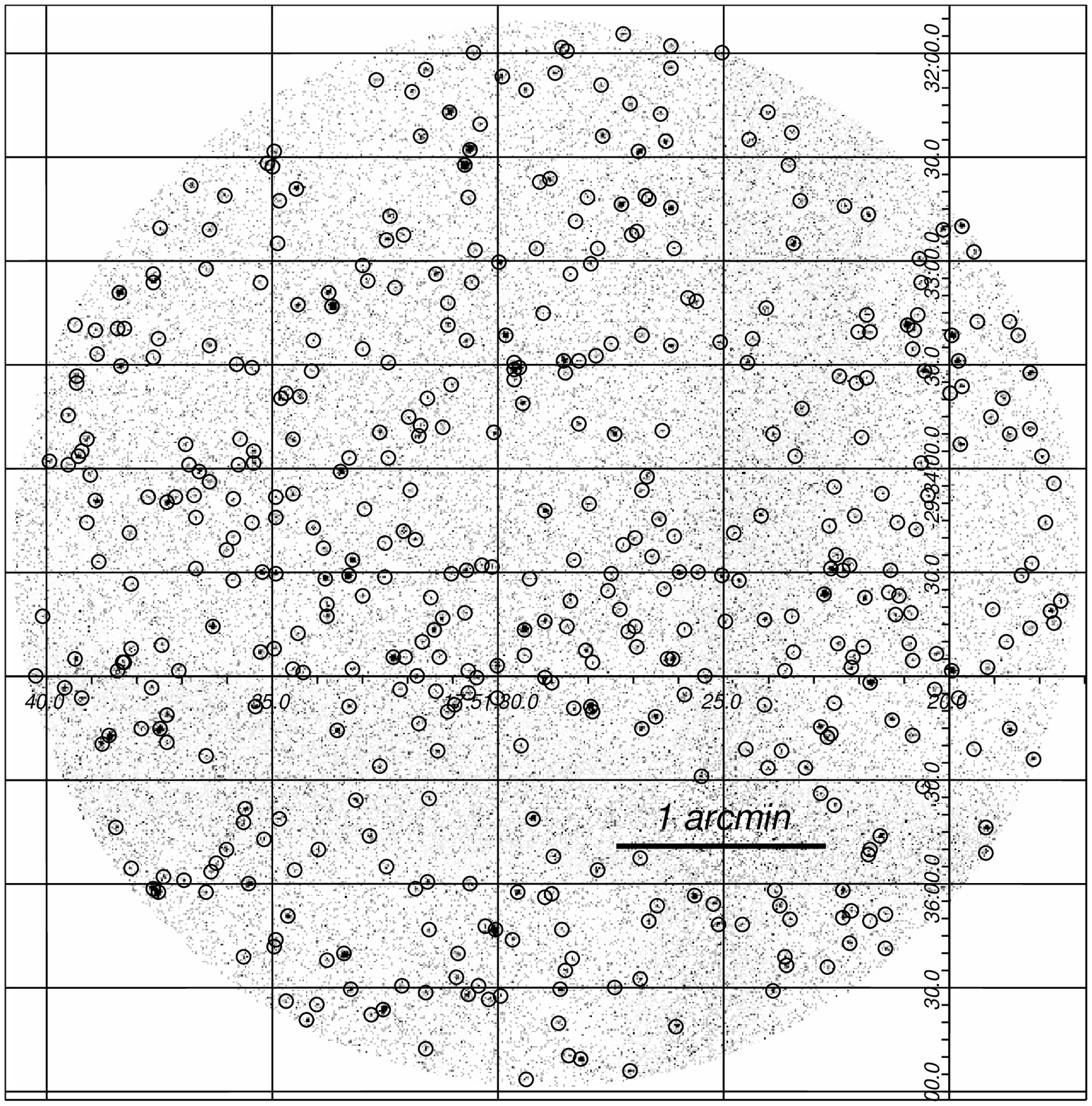} 
\caption{The deep {\sl Chandra} image of the studied region in 
energy band 0.5-7 keV. Circles of 2-arcsec radius denote the positions of
point sources detected after 1Msec exposure.  
The \emph{Chandra} data were reduced
following a standard procedure\cite{2005ApJ...628..655V}. The detector
background was modeled using the stowed dataset
(http://cxc.harvard.edu/contrib/maxim/stowed) and adjusted to the
conditions of the current observations using the count rate at
energies 9--12~keV, where \emph{Chandra} has virtually zero effective
area. The total measured X-ray surface brightness 
in this area is $I_{\rm
  3-7~keV}=(4.6\pm0.4)\times10^{-11}$~erg ~s$^{-1}$ ~cm$^{-2}$ ~deg$^{-2}$ 
in the 3--7~keV band, or equivalently $I_{\rm 2-10~keV}=(8.6\pm0.5)\times 
10^{-11}$ erg ~s$^{-1}$ ~cm$^{-2}$ ~deg$^{-2}$ in the 2--10~keV
band. Throughout the field there are noticeable  
variations of the soft X-ray ($<2$ keV) surface brightness due to what appears 
to be a perviously unknown supernova remnant shell projected onto the 
Chandra field. It should be noted that if a 1~Msec {\sl
  Chandra} observation were repeated in a nearby field, the measured
X-ray surface brightness would slightly differ because the number of
brightest point sources varies from field to field, an effect known
as cosmic variance in extragalactic studies. For the same reason,
there may be subtle field-to-field variations in the GRXE spectral shape, and in
particular in emission lines ratios, and recent observations indicated
that such variations to take place\cite{yamauchi08}. Additional 
variations of the spectrum of the unresolved Galactic X-ray emission
can be caused by the presence of genuine diffuse X-ray emitters
such as supernova remnants.
\label{image}
}
\end{figure}

\begin{figure}
\includegraphics[width=0.5\columnwidth]{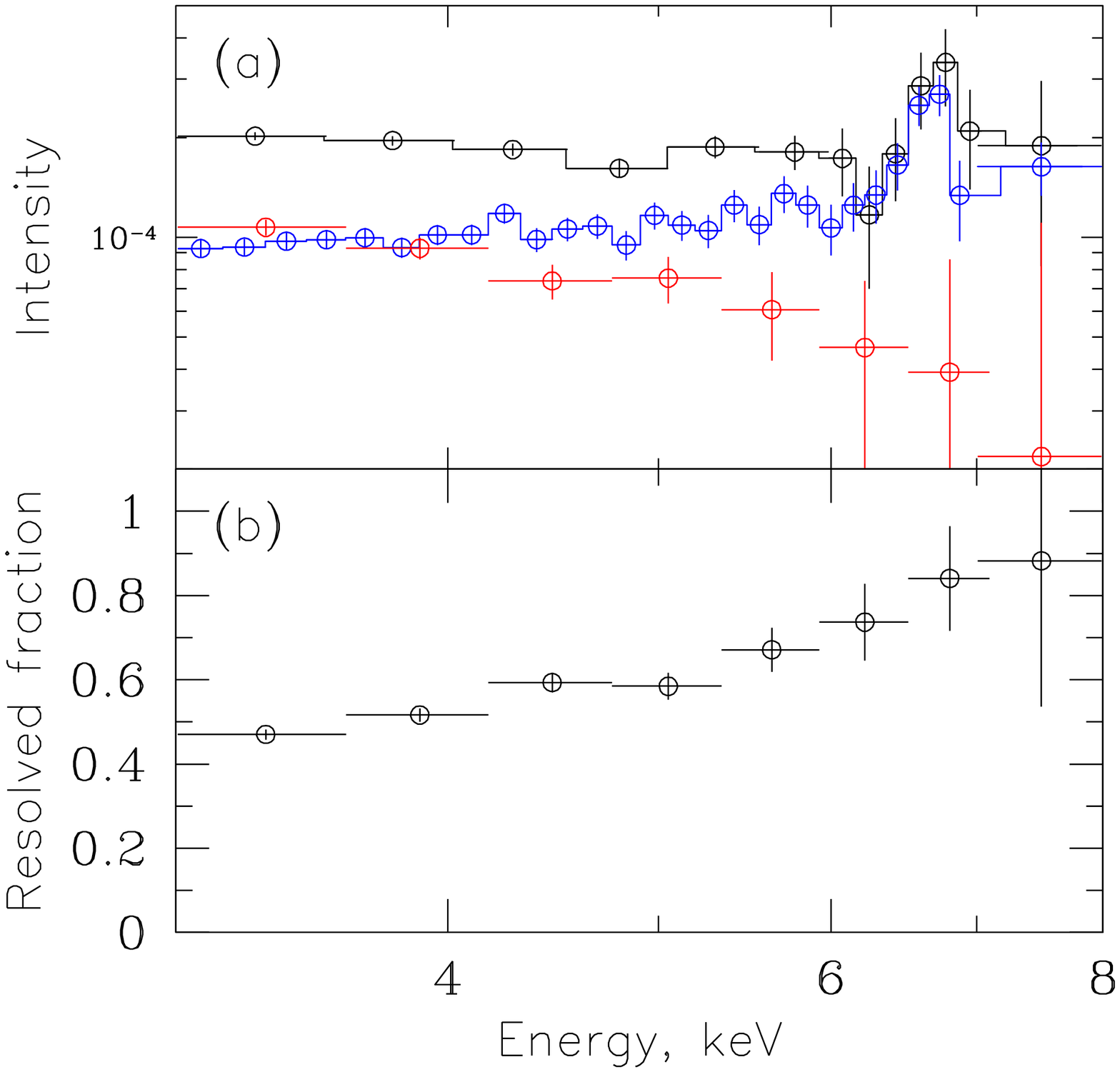}
\caption{GRXE spectrum and its resolved fraction. {\bf a):}
Spectra collected by {\sl Chandra} within the HRES region. Black
data points, error bars (68\% confidence intervals) and histogram show the spectrum of the total
emission from HRES; the collective spectrum of all detected sources
is presented in blue and the spectrum of the remaining unresolved
emission in the current observations is in red. The integrated
spectrum of detected sources exhibits a strong $\sim$6.7 keV iron
emission line, characteristic of hot (with temperatures 10--100
million K) plasma emission.  This line has been the main support for
the popular hypothesis that the GRXE has a truly diffuse, interstellar
origin, even though such hot interstellar plasma cannot be confined
within the Galaxy by its gravitational potential. Note that we took
into account that a small fraction of photons, $X$ (10\% at energies
4--6 keV, according to the {\em Chandra} Proposers' Observatory 
Guide\cite{chandra_pog}) from a point
source are scattered by the telescope outside the surrounding $2''$
radius circle. We therefore corrected the directly measured
collective spectrum of detected sources $F_1(E)$ through the formula
$\tilde{F_1}(E)=[F_1(E)-F_2(E)A_1/A_2]/[1-X-X A_1/A_2]$, where
$F_2(E)$ is the spectrum of the unresolved X-ray emission, $A_1$
($\sim 2$\% of the total) is the area covered by the $2''$ radius
circles used for collecting the source fluxes, and $A_2$ is the area
outside these circles. {\bf b):} Fraction of the X-ray
emission resolved by {\sl Chandra} into point sources as a function of
X-ray photon energy.
\label{spectra}
}
\end{figure}

We have detected sources in the broad 0.5--7~keV energy band in the
summed image of the HRES region (see Fig.~\ref{image}). The
sensitivity limit $f_{\rm lim}\sim 10^{-16}$ erg~s$^{-1}$~cm$^{-2}$
(minimum detectable flux in the 0.5--7 keV band corrected for the
interstellar absorption) corresponds to a minimum detectable
luminosity $L_{\rm 0.5-7~keV}\sim 10^{30}$ erg~s$^{-1}$ at a source
distance of 8 kpc, where most of the Galactic objects in this field
are expected to reside. In total, 473 sources have been detected with
statistical significance $>4\sigma$ (minimum number of counts per source
is about 10). In the upper panel of Fig.~\ref{spectra} we show the
energy spectrum of the total emission from HRES, as well as the two
components associated with the detected sources and with the remaining
unresolved emission. Most importantly, the summed spectrum of detected sources
exhibits a pronounced $\sim6.7$ keV iron emission line, a distinctive
feature of the GRXE which was often regarded as an important argument
in favour of it being the emission of a truly diffuse hot plasma
\cite{koyama86,tanaka02}. Only now we clearly see that 
the bulk of the 6.7~keV line emission, as well as of the neighbouring
continuum, is in fact produced by point sources. It is worth noting
that apart from the dominant 6.7 keV line, the unresolved (partially
due to finite energy resolution of the instrument and due to limited statistics
of the observation) blend of lines
at 6--7 keV may contain some contribution from 6.4 keV iron 
fluorescent emission, part of which may be unrelated to the GRXE and
result from irradiation of the interstellar medium by discrete 
X-ray sources\cite{sunyaev93,koyama96}.

The derived fraction of the X-ray emission resolved into point sources
is shown as a function of energy in the lower panel of Fig.~\ref{spectra}. 
{\em In the narrow energy band 6.5--7.1 keV containing the iron
emission line, $84\pm12$\% of the total X-ray emission is
resolved}. Moreover, we should recall again that the remaining
unresolved X-ray emission contains a non-negligible contribution from
the CXB. Assuming that the intensity of this unresolved
component in our 1~Msec {\em Chandra} observation is the same as in
the Chandra extragalactic deep fields\cite{hickox06} ($I_{\rm CXB,
  unresolved~1~Msec}=(3.4\pm1.7)\times10^{-12}$
erg~s~cm$^{-2}$~deg$^{-2}$ in the 2--8 keV energy band, or
$(2.9\pm1.4)\times10^{-13}$ erg~s~cm$^{-2}$~deg$^{-2}$ at 6.5--7.1
keV, assuming a power-law spectral shape with $\Gamma=1.4$), we can
estimate that $4\pm2$\% of the total intensity in the 6.5--7.1 keV band 
is unresolved CXB emission. {\em We conclude that we have resolved as much as
$88\pm12$\% of the GRXE emission into point sources at energies near
the 6.7~keV line}, the feature that was previously used as the
strongest argument in favour of a diffuse origin for the GRXE.

\begin{figure}
\includegraphics[width=0.5\columnwidth]{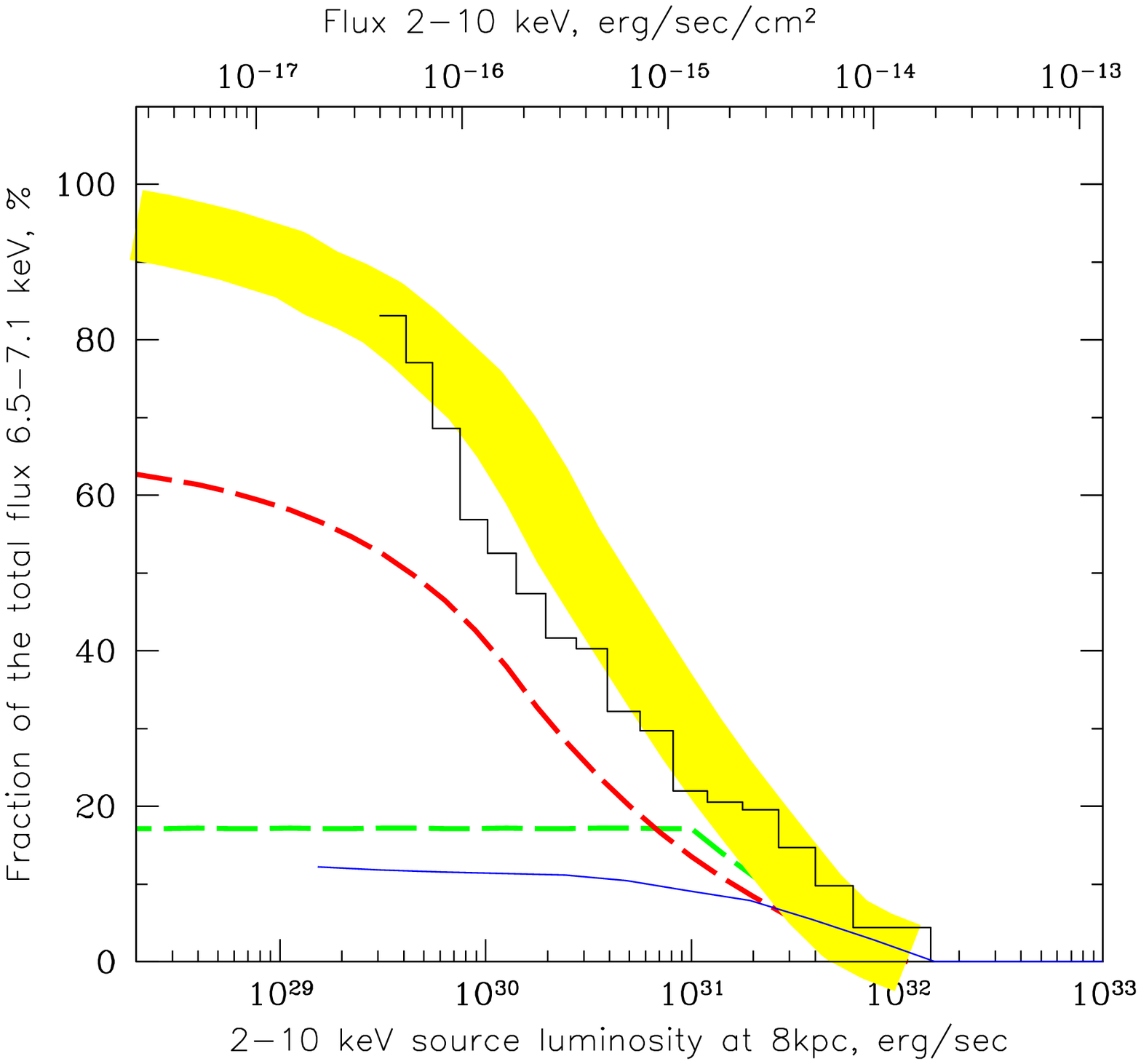}
\caption{Fraction of resolved X-ray emission around the 6.7~keV iron
emission line as a function of the limiting source
flux/luminosity. The histogram shows the fraction of the total flux in
the 6.5--7.1~keV energy band in the {\sl Chandra} field provided by
discrete sources with fluxes above a given detection threshold in the
2--10~keV energy band. The thick yellow curve shows the corresponding
dependence expected for a combination of Galactic sources with
luminosities below $10^{32}$ erg~s$^{-1}$ (all located at the Galactic
Centre distance of 8~kpc) and extragalactic sources with fluxes below
$2\times10^{-14}$ erg~s$^{-1}$~cm$^{-2}$ in this energy band. The blue
curve shows the expected contribution of extragalactic sources (mostly
active galactic nuclei) \cite{moretti03}. The green and
red curves show the expected contributions of accreting white dwarfs
and coronally active stars correspondingly (X-ray spectra of these
types of sources are described in e.g. \cite{hellier98,huenemoerder01}), 
estimated using the luminosity 
functions of these classes of objects measured in the Solar
vicinity\cite{sazonov06}. These Galactic curves were normalized so
that the total resolved fraction given by the model is equal to the
actual measured fraction of 84\% at the detection limit of the
1~Msec {\sl Chandra} observation of $10^{-16}$ erg~s$^{-1}$~cm$^{-2}$.
Fluxes of sources detected in the total {\em Chandra} band 0.5--7 keV
were converted into the 2--10 keV energy band using a count rate
dependent conversion factor that was estimated using stacked {\em
Chandra} spectra of bright, medium and faint sources ($>100$, 10--100
and 5--10 net counts in the image, respectively).  For Galactic
sources, their fluxes in the 2--10~keV band were converted to the
6.5--7.1 keV band using a single conversion factor estimated from the
stacked {\sl Chandra} spectrum of all detected sources. Comparison of
the stacked spectra of bright, medium and faint sources indicates that
this conversion factor does not vary by more than a factor of 1.3
around the adopted value. Conversion of fluxes of extragalactic
sources from 2--10 keV to 6.5--7.1 keV was done assuming that they
have power-law spectra with photon index $\Gamma=1.4$.
\label{res}
}
\end{figure}

Apart from a small contribution from extragalactic sources
(about 40--50 sources out of 473), most of the
sources detected by {\sl Chandra} in HRES are by all
likelihood accreting white dwarfs (with
luminosities $L_{\rm 2-10~keV}\sim 10^{31}$--$10^{32}$~erg~s$^{-1}$)
and binary stars with strong coronal activity (with $L_{\rm 2-10~keV}<
10^{31}$~erg~s$^{-1}$). Indeed, if we plot the fraction of the total GRXE flux
contained in sources with fluxes higher than a variable detection
threshold (see Fig.~\ref{res}), the resulting dependence proves to be
in good agreement with the expectation based on the luminosity
function of faint X-ray sources measured in the Solar
vicinity\cite{sazonov06}. Furthermore, since this locally
determined luminosity function continues to rise towards lower
luminosities, we can expect the still unresolved $\sim$10--20\% of the
GRXE flux also to be composed of coronally active and normal 
(Sun-like) stars with luminosities $L_{\rm 2-10~keV}<4\times10^{29}$
erg~s$^{-1}$, which are too weak to be detected at the Galactic Centre
distance even in ultra-deep {\sl Chandra} exposures. The contribution of
such faint stellar sources should rise with the decrease of the photon
energies because they typically have quite soft spectra. This might
be one of the reasons why we resolve more flux at high energies than
at lower energies.

The final resolution of the GRXE into discrete sources has
far-reaching consequences for our understanding of a variety of
astrophysical phenomena. Apart from the removal of a major energy
problem for the Galaxy, the important immediate outcome is that we can
now use the GRXE as a measure of the cumulative emission of faint
Galactic X-ray sources in the sense that spatial variations of the GRXE
properties over the Milky Way can indicate intrinsic variations in the
stellar populations. It has also become clear that the apparently diffuse X-ray
emission of external galaxies must contain, and in some cases be 
dominated by, unresolved emission from faint stellar-type
sources, namely accreting white dwarfs and coronally active stars.

\begin{addendum}
 \item MR thanks Maxim Markevitch for his help with the \emph{Chandra} 
instrumental background. This research was supported by the DFG cluster of 
excellence ``Origin and Structure of the Universe'', by NASA Chandra grant GO8-9132A, 
by program of Russian Academy of Sciences OFH-17, by grants RFFI 07-02-01004 
and RFFI 07-02-00961.

 \item[Correspondence] Correspondence and requests for materials
should be addressed to M. Revnivtsev
\end{addendum}

\newpage 

\epsfsize=0.9\textwidth
\epsfbox{./image_05_70_circles_2sec_gray.ps}

\epsfsize=0.9\textwidth
\epsfbox{./spectrum_gt3_res.ps}

\epsfsize=0.9\textwidth
\epsfbox{./cumcounts.ps}
\end{document}